\documentclass[conference,a4paper]{APSIPA2018}
\usepackage{multirow}

\usepackage{amsmath,graphicx,url,times}
\usepackage{amsfonts}

\usepackage{amsmath}
\usepackage[psamsfonts]{amssymb}
\usepackage{amsxtra}
\usepackage{threeparttable}

\usepackage{color}
\usepackage{url}
\usepackage{hyperref}

\usepackage{mathptmx}

\usepackage[margin=1in]{geometry}
\usepackage{amsmath,graphicx,amsfonts}
\usepackage[T1]{fontenc}
\usepackage{float}

\newcommand\sthanks[1]{%
  \begingroup
  \renewcommand\thefootnote{}\footnote{#1}%
  \addtocounter{footnote}{-1}%
  \endgroup
}

\begin{document}

\title{Performance Based Cost Functions for End-to-End Speech Separation}

\author{%
\authorblockN{%
Shrikant Venkataramani\authorrefmark{1}\authorrefmark{2},
Ryley Higa\authorrefmark{1}\authorrefmark{2},
Paris Smaragdis\authorrefmark{2}\authorrefmark{3} 
}
\authorblockA{%
\authorrefmark{2}
University of Illinois at Urbana-Champaign \\
\authorblockA{%
\authorrefmark{3}
Adobe Research}
\authorblockA{%
\authorrefmark{1}
Authors with equal contribution}
\{svnktrm2, higa2\}@illinois.edu}

}

\maketitle
\thispagestyle{empty}

\begin{abstract}
Recent neural network strategies for source separation attempt to model audio signals by processing their waveforms directly. Mean squared error (MSE) that measures the Euclidean distance between waveforms of denoised speech and the ground-truth speech, has been a natural cost-function for these approaches. However, MSE is not a perceptually motivated measure and may result in large perceptual discrepancies. In this paper, we propose and experiment with new loss functions for end-to-end source separation. These loss functions are motivated by BSS\_Eval and perceptual metrics like source to distortion ratio (SDR), source to interference ratio (SIR), source to artifact ratio (SAR) and short-time objective intelligibility ratio (STOI). This enables  the flexibility to mix and match these loss functions depending upon the requirements of the task. Subjective listening tests reveal that combinations of the proposed cost functions help achieve superior separation performance as compared to stand-alone MSE and SDR costs.\\
\end{abstract}


\sthanks{This work was supported by NSF grant 1453104}

\section{Introduction}
Single channel source separation deals with the problem of extracting the speaker or sound of interest from a mixture consisting of multiple simultaneous speakers or audio sources. In order to identify the source, we assume the availability of a few unmixed training examples. These examples are used to build representative models for the corresponding source. With the development of deep learning, several neural network architectures have been proposed to solve the supervised single-channel source separation problem~\cite{isik2016single, chen2017deep, luo2018speaker}. The latest deep-learning approaches to source separation have started to focus on performing separation by operating directly on the mixture waveforms~\cite{venkataramani_adaptive_2017, luo2018tasnet, Fu2017Raw, rethage2018wavenet}. To train these end-to-end models, the papers restrict themselves to minimizing a mean-squared error loss~\cite{venkataramani_adaptive_2017, luo2018tasnet, Fu2017Raw}, an L1 loss~\cite{rethage2018wavenet} or a source-to-distortion ratio~\cite{venkataramani_adaptive_2017, luo2018tasnet} based cost-function between the separated speech and the corresponding ground-truth. A potential direction for improving end-to-end models is to use loss functions that capture the salient aspects of source separation. Predominantly, the BSS\_Eval metrics source-to-Distortion ratio~(SDR), source-to-Interference ratio~(SIR), source-to-Artifact ratio~(SAR)~\cite{fevotte2005bss_eval}, and short-time objective intelligibility~(STOI)~\cite{taal2010short} have been used to evaluate the performance of source separation algorithms. Fu et.al., have proposed an end-to-end neural network that captures the effect of STOI in performing source separation~\cite{fu2017end}. Alternatively, we could also develop suitable cost-functions for end-to-end source separation by interpreting these metrics as suitable loss functions themselves. Proposing and evaluating these new cost-functions for source separation would also allow us mix and match a combination of these metrics to suit our requirements and improve source separation performance, for any neural network architecture.\\

Section~\ref{sec:endtoend} provides a description of the neural network used for end-to-end source separation. Section~\ref{sec:costfunctions} presents the approach to interpret the BSS\_Eval and STOI metrics as loss functions for end-to-end source separation. We evaluate our cost functions by deploying subjective listening tests. The details of our experiments, subjective listening tests and the corresponding results are discussed in section~\ref{sec:experiments} and we conclude in section~\ref{sec:conclusion}. \\

\begin{figure}[t!]
\centering
  \includegraphics[width=\columnwidth]{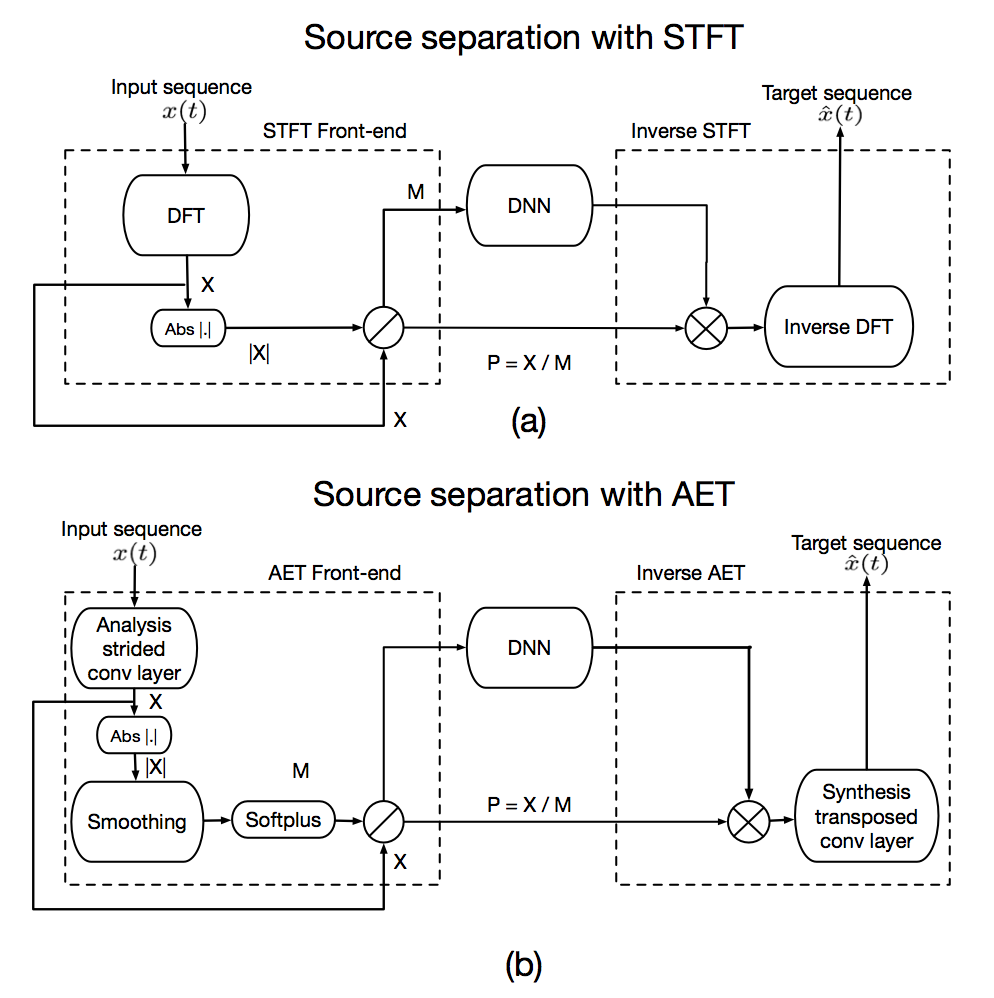}
  \caption{(a) Block diagram of a source separation network using STFT as the front-end. (b) Block diagram of the equivalent end-to-end source separation network using an auto-encoder transform as the front-end.}~\label{fig:STFTAET}
\end{figure}

\begin{figure*}[ht!]
\centering
  \includegraphics[width=\textwidth]{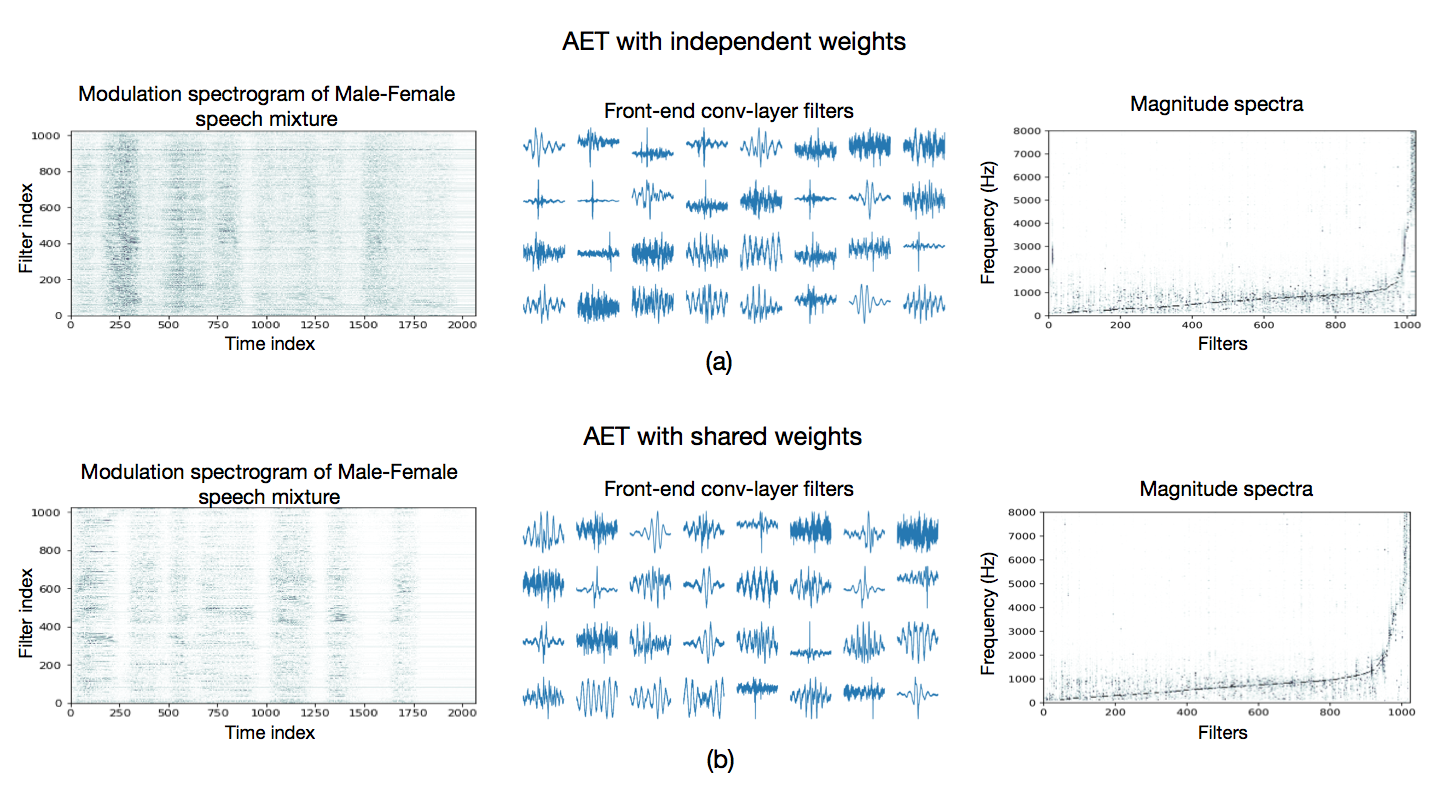}
  \caption{(a) An example of the modulation spectrogram for a male-female speech mixture (left), adaptive bases i.e., filters of analysis convolutional layer (middle), Normalized magnitude spectra of adaptive bases (right) for independent analysis and synthesis layers (top). (b) An example of the modulation spectrogram for a male-female speech mixture (left), adaptive bases i.e., filters of analysis convolutional layer (middle), Normalized magnitude spectra of adaptive bases (right) for shared analysis and synthesis layers (bottom). The orthogonal-AET uses a transposed version of the analysis filters for the synthesis convolutional layer. The filters are ordered according to their dominant frequency component (from low to high). In the middle subplots, we show a subset of the first $32$ filters. The adaptive bases concentrate on the lower frequencies and spread-out at the higher frequencies. These plots have been obtained using SDR as the cost function.}~\label{fig:AETbases}
\end{figure*}
\vspace{-3mm}

\section{End-to-end separation}
\label{sec:endtoend}
We first describe the end-to-end neural network architecture used for source separation. We begin with the description of a short-time Fourier transform~(STFT) based source separation neural network. This network can be transformed into an end-to-end separation network by replacing the STFT analysis and synthesis operations by their neural network alternatives~\cite{venkataramani_adaptive_2017}. \\

Figure~\ref{fig:STFTAET} (a) shows the architecture of a source separation network~\cite{venkataramani_adaptive_2017}. The flow of data through the network can be explained by the following sequence of steps. The mixture is first transformed into its equivalent time-frequency (TF) representation using the STFT. The TF representation is then split into its magnitude and phase components. The magnitude spectrogram of the mixture is then fed to the separation neural network. This network is trained to estimate the magnitude spectrogram of the source of interest from the magnitude spectrogram of the mixture. The estimated magnitude spectrogram is multipled by the phase of the mixture and transformed into the time domain by the overlap-and-add approach to invert the STFT.

As described in~\cite{venkataramani_adaptive_2017}, we can transform this network into an end-to-end source separation network by replacing the STFT blocks by corresponding neural networks, with the following sequence of steps. (i) The STFT and inverse STFT operations can be replaced by 1-D convolution and transposed convolution layers. This would enable the network to learn an adaptive TF representation~($\mathbf{X}$) directly from the waveform of the mixture. (ii) The front-end convolutional layer needs to be followed by a smoothing convolutional layer. This is done to obtain a smooth modulation spectrogram~($\mathbf{M}$) that is similar to STFT magnitude spectrogram. The carrier component obtained using the element-wise division operation, ~$\mathbf{P} = \mathbf{X}/\mathbf{M}$ incorporates the rapid variations of the adaptive TF representation. We will refer to this front end as the auto-encoder transform~(AET). Figure~\ref{fig:STFTAET}(b) gives the block diagram of the end-to-end separation network using an AET front-end.


\subsection{Examining the adaptive bases}
\label{ssec:aetbases}
We can understand the performance of end-to-end source separation better by examining the learned TF bases and TF representations. Figure~\ref{fig:AETbases} plots the modulation spectrograms of a male-female speech mixture, the first $32$ TF bases and their corresponding magnitude spectra. We rank the TF bases according to their dominant frequency component. We give these plots for two cases viz., the analysis convolution and synthesis transposed-convolution layers are independent (top), the analysis convolution and synthesis transposed-convolution layers share their weights (bottom). We observe that, similar to STFT bases, the adaptive bases are frequency selective in nature. However, the adaptive bases are concentrated at the lower frequencies and spread-out at the higher frequencies similar to the filters of the Mel filter bank.


\section{Performance based cost-functions}
\label{sec:costfunctions}
Source separation approaches have traditionally relied on the use of magnitude spectrograms as the choice of TF representation. Magnitude spectrograms have been interpreted as probability distribution functions~(pdf) drawn from random variables of varying characteristics. This motivated the use of several cost functions like the mean squared error~\cite{lee2001algorithms}, Kullback-Leibler divergence~\cite{lee2001algorithms}, Itakura-Saito divergence~\cite{fevotte2009nonnegative}, Bregman divergences~\cite{sra2006generalized} to be used for source separation. Since these interpretations do not extend to waveforms, there is a need to propose and experiment with additional cost-functions suitable for use in the waveform domain. As stated before, the BSS\_Eval metrics (SDR, SIR, SAR) and STOI are the most commonly used metrics to evaluate the performance of source separation algorithms. We now discuss how we can interpret these metrics as suitable loss functions for our neural network. 

\subsection{BSS\_Eval based cost-functions}
In the absence of external noise, the distortions present in the output of a source separation algorithm can be categorized as interference and artifacts. Interference refers to the lingering effects of the other sources on the separated source. Thus, source-to-interference ratio (SIR) is a metric that captures the ability of the algorithm to eliminate the other sources and preserve the source of interest. The processing steps in an algorithm may introduce arifacts or additional sounds in the separation results that do not exist in the original sources. Source-to-artifact ratio (SAR) measures the ability of the network to produce high quality results without introducing additional artifacts. The unwanted non-linear processing effects that may occur due to a neural network are also incorporated by SAR. These metrics can be combined into source-to-distortion ratio (SDR), which captures the overall separation quality of the algorithm. We denote the output of the network by $\mathbf{x}$. This output should ideally be equal to the target source $\mathbf{y}$ and completely suppress the interfering source~$\mathbf{z}$. We note that notations refer to the time-domain waveforms of each signal. Thus, $\mathbf{y}$ and $\mathbf{z}$ are constants with respect to any optimization (max or min) applied on the network output $\mathbf{x}$. We will also use the following definition of the inner-product between vectors as,  $\langle \mathbf{xy}\rangle = \mathbf{x}^{T} \cdot \mathbf{y}$  

Maximizing SDR with respect to $\mathbf{x}$ can be given as,   
 \label{ssec:bssevalcostfunctions}
 \begin{align*}
     \max \text{SDR}(\mathbf{x}, \mathbf{y}) &= \max \frac{\langle \mathbf{xy}\rangle^2}{\langle \mathbf{y y} \rangle \langle \mathbf{x x} \rangle - \langle \mathbf{xy} \rangle^2} \\
     & \equiv \min \frac{\langle \mathbf{y y} \rangle \langle \mathbf{x x} \rangle - \langle \mathbf{xy} \rangle^2}{\langle \mathbf{xy}\rangle^2} \\
     & = \min \frac{\langle \mathbf{yy} \rangle \langle \mathbf{xx} \rangle}{\langle \mathbf{xy} \rangle^2} - \frac{\langle \mathbf{xy} \rangle^2}{\langle \mathbf{xy} \rangle^2} \\
     & \propto \min \frac{\langle \mathbf{xx} \rangle}{\langle \mathbf{xy} \rangle^2}
 \end{align*}

Thus, maximizing the SDR is equivalent to maximizing the correlation between $\mathbf{x}$ and $\mathbf{y}$, while producing the solution with least energy. Maximizing the SIR cost function can be given as, 

 \begin{align*}
    \max \text{SIR}(\mathbf{x}, \mathbf{y}, \mathbf{z}) &= \max \frac{\langle \mathbf{zz} \rangle^2 \langle \mathbf{xy}\rangle^2}{\langle \mathbf{yy} \rangle^2 \langle \mathbf{xz} \rangle^2}
    \equiv \min \frac{\langle \mathbf{xz} \rangle^2}{ \langle \mathbf{xy}\rangle^2}
 \end{align*}
 
 Maximizing SIR is equivalent to maximizing the correlation between the network output $\mathbf{x}$ and target source $\mathbf{y}$ while minimizing the correlation between $\mathbf{x}$ and interference $\mathbf{z}$. Over informal listening tests, we identified that a network trained purely on SIR, maximizes time-frequency (TF) bins where the target is present and the interference is not present and minimizes TF bins where both sources are present or bins where the interference dominates the target. This results in a network output consisting of sinusoidal tones near TF bins dominated by the target source. \\   
 
 For the SAR cost function, we assume that the clean target source $\mathbf{y}$ and the clean interference $\mathbf{z}$ are orthogonal in time. This allows for the following simplification:  
 \begin{align*}
     \max \text{SAR}(\mathbf{x}, \mathbf{y}, \mathbf{z}) &= \max \frac{\lVert \frac{\langle \mathbf{xy} \rangle}{\langle \mathbf{yy} \rangle} \mathbf{y} + \frac{\langle \mathbf{xz} \rangle}{\langle \mathbf{zz} \rangle} \mathbf{z} \rVert^2} {\lVert \mathbf{x} - \frac{\langle \mathbf{xy} \rangle}{\langle \mathbf{yy} \rangle} \mathbf{y} - \frac{\langle \mathbf{xz} \rangle}{\langle \mathbf{zz} \rangle} \mathbf{z} \rVert^2} \\
     & \equiv \min \frac{\lVert \mathbf{x} - \frac{\langle \mathbf{xy} \rangle}{\langle \mathbf{yy} \rangle} \mathbf{y} - \frac{\langle \mathbf{xz} \rangle}{\langle \mathbf{zz} \rangle} \mathbf{z} \rVert^2}
     {\lVert \frac{\langle \mathbf{xy} \rangle}{\langle \mathbf{yy} \rangle} \mathbf{y} + \frac{\langle \mathbf{xz} \rangle}{\langle \mathbf{zz} \rangle} \mathbf{z} \rVert^2} \\
     & = \min \frac{\langle \mathbf{xx} \rangle - \frac{\langle \mathbf{xy} \rangle^2}{\langle \mathbf{yy} \rangle} - \frac{\langle \mathbf{xz} \rangle^2}{\langle \mathbf{zz} \rangle}}
     {\frac{\langle \mathbf{xy} \rangle^2}{\langle \mathbf{yy} \rangle} + \frac{\langle \mathbf{xz} \rangle^2}{\langle \mathbf{zz} \rangle}} \\
     & \propto \min \frac{\langle \mathbf{xx} \rangle}
     {\frac{\langle \mathbf{xy} \rangle^2}{\langle \mathbf{yy} \rangle} + \frac{\langle \mathbf{xz} \rangle^2}{\langle \mathbf{zz} \rangle}}
 \end{align*}
 
 From the equations, we see that SAR does not distinguish between the target source and the interference. Consequently, optimizing the SAR cost function does not directly optimize the quality of separation. The purpose of optimizing the SAR cost function should be to reduce audio artifacts in conjunction with a loss function that penalizes the presence of interference such as the SIR. In practice, a network that optimizes SAR directly should apply the identity transformation to the input mixture.

\subsection{STOI based cost function}
 \label{ssec:stoicostfunction}
 The drawback of BSS\_Eval metrics is that they fail to incorporate ``intelligibility'' of the separated signal. Short-time objective intelligibility (STOI)~\cite{taal2010short} is a metric that correlates well with subjective speech intelligibility. STOI accesses the short-time correlation between TF representations of target speech~$\mathbf{y}$ and network output~$\mathbf{x}$. We now describe the sequence of steps involved in interpreting STOI as a cost-function. 
 
 The network output $\mathbf{x}$ and target source~$\mathbf{y}$ waveforms are first transformed into the TF domain using an STFT step. To do so, we use Hanning windowed frames of $256$ samples zero-padded to a size of $512$ samples each, and a hop of $50$\%. This STFT step was implemented using a $1$-D convolution operation. The resulting magnitude spectrograms are transformed into an octave-band representation by grouping frequencies using $15$ one-third octave bands reaching upto $10000$ Hz. The resulting representations will be denoted as $\mathbf{\hat{X}}$ and $\mathbf{\hat{Y}}$, corresponding to $\mathbf{x}$ and $\mathbf{y}$ respectively. The representation $\mathbf{\hat{X}_{j, m}}$ corresponds to the $j$th one-third octave band at the $m$th time frame. This was implemented as a matrix multiplication step applied on the magnitude spectrograms. 
 
 Given the one-third octave band representation $\mathbf{\hat{X}}$ and $\mathbf{\hat{Y}}$, we constructed new vectors $\mathbf{X_{j, m}}$ and $\mathbf{Y_{j, m}}$ consisting of $N = 30$ previous frames before the $m$th time frame. We can write this explicitly as 
 \begin{align*}
     \mathbf{X_{j, m}} = [\mathbf{\hat{X}_{j, m - N + 1}}, \mathbf{\hat{X}_{j, m - N + 2}}, ..., \mathbf{\hat{X}_{j, m}}]^T
 \end{align*}
 
 Let $\mathbf{X_{j, m}(n)}$ be the $n$th frame of vector $\mathbf{X_{j, m}}$. The octave-band representation of the network output is then normalized and clipped to have the similar scale as the target source, which is denoted as $\mathbf{\bar{X_{j, m}}}$. The clipping procedure   clips the network output so that the signal-to-distortion ratio (SDR) is above $\beta = -15dB$.  
 \begin{align*}
     \mathbf{\bar{X}_{j, m}}(n) = \min \left (\frac{\lVert \mathbf{Y_{j, m}} \rVert}{\lVert \mathbf{X_{j, m}} \rVert} \mathbf{X_{j, m}}(n), (1 + 10^{-\beta/20}) \mathbf{Y_{j, m}}(n) \right) 
 \end{align*}
 
  We then compute the intermediate intelligibility matrix denoted by $d_{j, m}$ by taking the correlation between $\mathbf{\bar{X}_{j,m}}$ and $\mathbf{Y_{j,m}}$.  
 \begin{align*}
     d_{j, m} = \frac{(\mathbf{\bar{X}_{j, m}} - \mathbf{\mu_{\bar{X}_{j, m}}})^T(\mathbf{Y_{j, m}} - \mathbf{\mu_{Y_{j, m}}})}
     {\lVert \mathbf{\bar{X}_{j, m}} - \mathbf{\mu_{\bar{X}_{j, m}}} \rVert \lVert \mathbf{Y_{j, m}} - \mathbf{\mu_{Y_{j, m}}} \rVert}
 \end{align*}
 
 To get the final STOI cost function, we take the average short-time correlation over $M$ total time frames and $J = 13$ total one-third octave bands.  
 \begin{align*}
     \text{STOI} = \frac{1}{JM} \sum_{j, m} d_{j, m}
 \end{align*}
 It is clear by the procedure that maximizing the STOI cost function is equivalent to maximizing the average short-time correlation between the TF representations for the target source and separation network output.

\begin{figure*}[t!]
\centering
  \includegraphics[trim={1cm 1cm 1cm 1cm}, width=\textwidth]{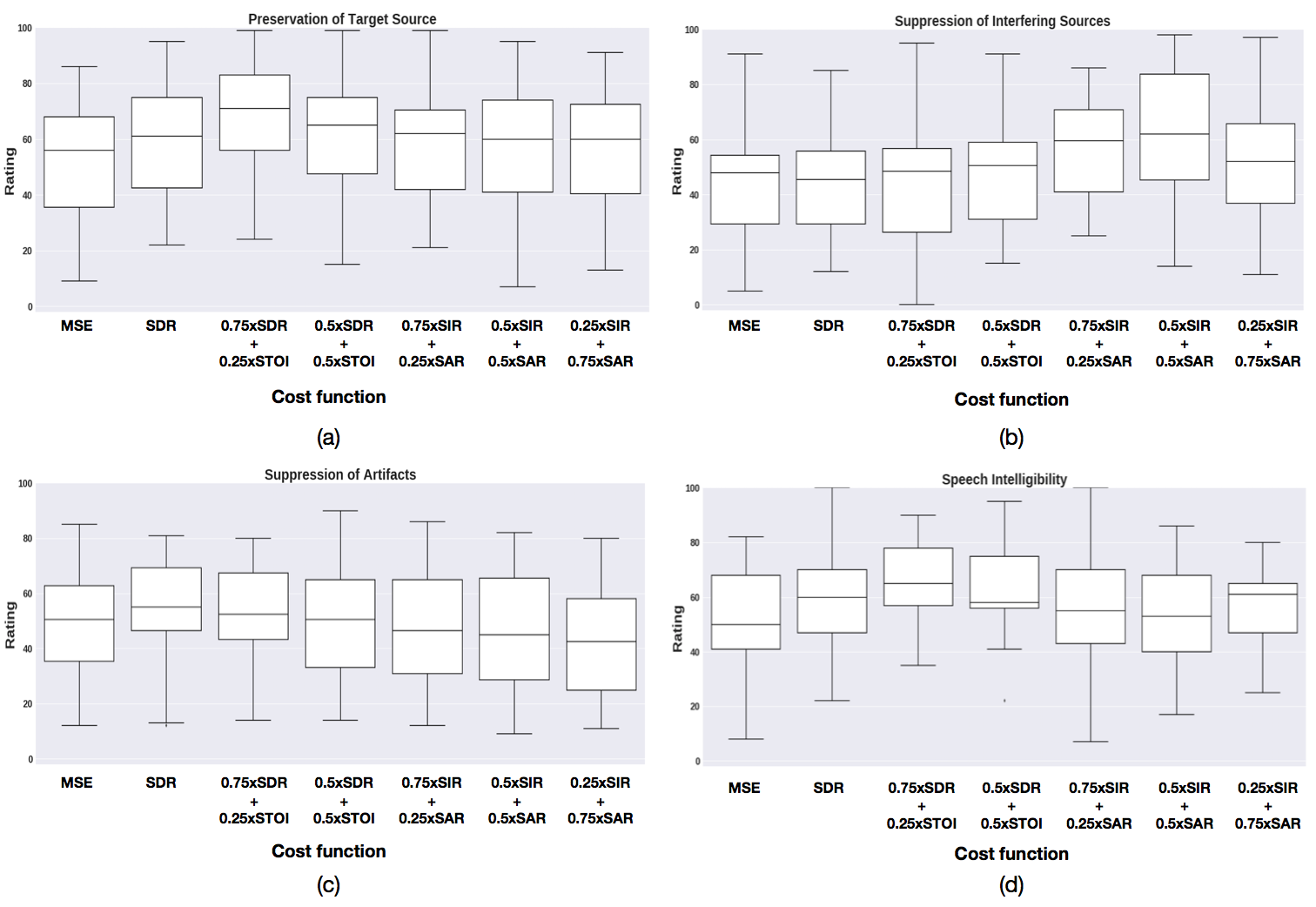}
  \caption{ Listening test scores for tasks (a) Preservation of target source. (b) Suppression of interfering sources. (c) Suppression of artifacts (d) Speech intelligibility over different cost-functions. The distribution of scores is presented in the form of a box-plot where, the solid line in the middle give the median value and the extremities of the box give the $25^{\text{th}}$ and $75^{\text{th}}$ percentile values.}~\label{fig:MTurk_results}
\end{figure*}

\section{Experiments}
\label{sec:experiments}
Since the paper deals with interpreting source separation metrics as a cost function, it is not a reasonable approach to use the same metrics for their evaluation. In this paper, we use subjective listening tests targeted at evaluating the separation, artifacts and intelligibility of the separation results to compare the different loss functions. We use the crowd-sourced audio quality evaluation (CAQE) toolkit~\cite{cartwright_fast_2016} to setup the listening tests over Amazon Mechanical Turk~(AMT). The details and results of our experiments follow. 

\subsection{Experimental Setup}
\label{ssec:experimentalsetup}
For our experiments, we use the end-to-end network shown in figure~\ref{fig:STFTAET}(b). The separation was performed with a $1024$ dimensional AET representation computed at a stride of $16$ samples. A smoothing of $5$ samples was applied by the smoothing convolutional layer. The separation network consisted of $2$ dense layers each followed by a softplus non-linarity. This network was trained using different proposed cost functions and their combinations. We compare the cost-functions by evaluating their performance on isolating the female speaker from a mixture comprising a male speaker and a female speaker, using the above end-to-end network.

To train the network, we randomly selected $15$ male-female speaker pairs from the TIMIT database~\cite{timit}. $10$ pairs were used for training and the remaining $5$ pairs were used for testing. Each speaker has $10$ recorded sentences in the database. For each pair, the recordings were mixed at $0$ dB. Thus, the training data consisted of $100$ mixtures. The trained networks were compared on their separation performance on the $50$ test sentences. Clearly, the test speakers were not a part of the training data to ensure that the network learns to separate female speech from a mixture of male and female speakers and does not memorize the speakers themselves. 

In the subjective listening tests we compare the performance of end-to-end source separation under the following cost functions: \\
(i) Mean squared error \\
(ii) $SDR$ \\
(iii) $0.75 \times SDR + 0.25 \times STOI$ \\
(iv) $0.5 \times SDR + 0.5 \times STOI$ \\
(v) $0.75 \times SIR + 0.25 \times SAR$ \\
(vi) $0.5 \times SIR + 0.5 \times SAR$ \\
(vii) $0.25 \times SIR + 0.75 \times SAR$. \\
These combinations were selected to understand the effects of individual cost functions on separation performance. We scale the value of each cost-function to unity before starting the training procedure. This was done to control the weighting of terms in case of composite cost-functions.

\subsection{Evaluation}
\label{ssec:evaluation}
Using CAQE over a web environment like AMT has been shown to give consistent results to listening tests performed in controlled lab environments~\cite{cartwright_fast_2016}. Thus, we use the same approach for our listening tests. The details are briefly described below.

\subsubsection{Recruiting Listeners}
\label{sssec:recruiting}
For the listening tasks, we recruited listeners on Amazon Mechanical Turk that were over the age of 18 and had no previous history of hearing impairment. Each listener had to pass a brief hearing test that consisted of identifying the number of sinusoidal tones within two segments of audio. If the listener failed to identify the correct number of tones within the audio clip in two attempts, the listener's response was rejected. For the listening tests, we recruited a total of $180$ participants over AMT. 

\subsubsection{Subjective Listening Tests}
\label{sssec:listeningtests}
We assigned each of the accepted listeners to one of four evaluation tasks. Each task asked listeners to rate the quality of separation based on one of four perceptual metrics: preservation of source, suppression of interference, absence of additional artifacts, and speech intelligibility. The perceptual metrics such as preservation of source, suppression of interference, absence of additional artifacts, and speech intelligibility directly correspond to objective metrics such as SDR, SIR, SAR, and STOI respectively.

Accepted listeners were given the option to submit multiple evaluations for each of the different tasks. For each task, we trained listeners by giving each listener an audio sample of the isolated target source as well as a mixture of the source and interfering speech. We also provided 1-3 audio separation examples of poor quality and 1-3 audio examples of high quality according to the perceptual metric assigned to the listener. The audio files used to train the listener all had exceptionally high or low objective metrics (SDR, SIR, SAR, STOI) with respect to the pertaining task so that listeners could base their ratings in comparison to the best or worst separation examples.

After training, the listeners were then asked to rate eight unlabelled, randomly ordered, separation samples from 0 to 100 based on the metric assigned. The isolated target source was included in the listener evaluation as a baseline. The other seven audio samples correspond to separation examples output by a neural network trained with different cost functions enlisted in section~\ref{ssec:experimentalsetup}. 

\subsection{Results and Discussion}
\label{ssec:results}
Figure~\ref{fig:MTurk_results} gives the results of the subjective listening tests performed through AMT for each of the four tasks. The results are shown in the form of a bar-plot that shows the median value (solid line in the middle) and the $25$-percentile and $75$-percentile points (box boundaries). The vertical axis gives the distribution of listener-scores over the range (0-100) obtained from the tests. The horizontal axis shows the different cost-functions used for evaluation, as listed in section~\ref{ssec:experimentalsetup}. This also helps us to understand the nature of the proposed cost-functions. For example, figure \ref{fig:MTurk_results}(b) (bars 5,6,7) shows that incorporating the SIR term into the cost function explicitly, helps the network to suppress the interfering sources better. Similarly, the addition of a STOI term into the cost function improves the results in terms of speech intelligibility as seen in figure \ref{fig:MTurk_results}(d). It is also observed that adding STOI to the SDR cost-function helps in preserving the target source better (figure \ref{fig:MTurk_results}(a), bars 2,3 and 4). One possible reason for this could be that increasing the intelligibility of the separation results results in a perceptual notion of preserving the target source better. The BSS\_Eval cost functions appear to be comparable in terms of preserving the target source (figure \ref{fig:MTurk_results}(a), bars 2,5,6,7) and slightly better than MSE. In terms of artifacts in the separated source, SDR outperforms all the cost-functions, all of which seem to introduce a comparable level of artifacts into the separation results (figure \ref{fig:MTurk_results}(c)). The use of SAR in the cost-function does not seem to have favorable or adverse effects on the perception of artifacts on the separation results.\\

\section{Conclusion and Future Work}
\label{sec:conclusion}
In this paper we have proposed and experimented with novel cost-functions motivated by BSS\_Eval and STOI metrics,  for end-to-end source separation. We have shown that these cost-functions capture different salient aspects of source-separation depending upon their characteristics. This enables the flexibility to use composite cost-functions that can potentially improve the performance of existing source separation algorithms. \\

\bibliographystyle{IEEEtran}
\bibliography{refs}

\end{document}